\documentclass[11pt]{article}
\usepackage{graphicx}
\usepackage[margin=1.25in]{geometry}
\usepackage[usenames,dvipsnames]{color}
\usepackage{url}
\usepackage[colorlinks = true,
            linkcolor = blue,
            urlcolor  = blue,
            citecolor = blue,
            anchorcolor = blue]{hyperref}
\usepackage{authblk}

\def\beq{\begin{equation}}
\def\eeq#1{\label{#1}\end{equation}}
\def\eeqn{\end{equation}}

\newenvironment{Eqnarray}%
   {\arraycolsep 0.14em\begin{eqnarray}}{\end{eqnarray}}
\def\beqa{\begin{Eqnarray}}
\def\eeqa#1{\label{#1}\end{Eqnarray}}
\def\eeqan{\end{Eqnarray}}

\let\bar=\overbar

\def\lsim{\mathrel{\raise.3ex\hbox{$<$\kern-.75em\lower1ex\hbox{$\sim$}}}}
\def\gsim{\mathrel{\raise.3ex\hbox{$>$\kern-.75em\lower1ex\hbox{$\sim$}}}}

\def\del{\partial}
\def\Dslash{\not{\hbox{\kern-4pt $D$}}}
\def\dslash{\not{\hbox{\kern-2pt $\del$}}}
\def\pslash{\not{\hbox{\kern-2pt $p$}}}
\def\ETmiss{\not{\hbox{\kern-4pt $E$}}_T}

\def\Dlr{\mathrel{\raise1.5ex\hbox{$\leftrightarrow$\kern-1em\lower1.5ex\hbox{$D$}}}}

\def\MSB{{\bar{M \kern -2pt S}}}
\def\msb{{\bar{\scriptsize M \kern -1pt S}}}

\def\drb{{\bar{\scriptsize D \kern -1pt R}}}

\newcommand\snowmass{\begin{center}\rule[-0.2in]{\hsize}{0.01in}\\\rule{\hsize}{0.01in}\\
\vskip 0.1in Submitted to the  Proceedings of the US Community Study\\ 
on the Future of Particle Physics (Snowmass 2021)\\ 
\rule{\hsize}{0.01in}\\\rule[+0.2in]{\hsize}{0.01in} \end{center}}

\author[1]{Julie Hogan}
\author[2,3]{Aneliya Karadzhinova-Ferrer}
\author[4]{Sudhir Malik}

\affil[1]{Department of Physics \& Engineering, Bethel University, St.~Paul MN, 55112, USA} 
\affil[2]{Helsinki Institute of Physics, P.O. Box 64, 00014, University of Helsinki, Finland}
\affil[3]{Lappeenranta University of Technology, P.O. Box 20, 53851 Lappeenranta, Finland}
\affil[4]{Department of Physics, University of Puerto Rico, Mayagüez PR, 00681, USA}

\title{Summary Report of the Topical Group on \\ Career Pipeline and Development (CommF2)\\
Community Engagement Frontier\\
Snowmass 2021\\}

\begin{document}
\maketitle



 \begin{abstract}
The HEP faculty hire job market has stayed fairly plateaued over the years not keeping up with number of postdocs and PhD produced seeking such employment. Physicists who seek jobs outside this realm face challenges. For example those hired as faculties at predominantly undergraduate institutions (PUI) and community colleges (CC) face hurdles to keep with research due higher teaching load and funding challenges.  At the same time those who seek employment in industry may find themselves in unprepared territory despite marketable skills. New job opportunities seeking HEP developed skills have appeared in data science, machine learning and quantum computing. 
Given that a vast majority transition to the industry jobs, we must strengthen the existing paths for this transition and develop new ways to facilitating it. A strong engagement between HEP and its alumni would boost this process. At the same time those who stay in academia but choose to work at PUIs or CCs must be enabled to continue to pursue research and receive support and guidance for funding. PUIs and CCs serve as a gateway to opportunities for inclusiveness beyond national labs and academic research institutions offering an early starting point in the pipeline that can mitigate issues of lack of diversity and underrepresented participation of different groups in HEP. This report summarises the study undertaken to investigate these issues in HEP community and provide findings and recommendations.

\snowmass

\end{abstract}

\newpage

\def\thefootnote{\fnsymbol{footnote}}
\setcounter{footnote}{0}

\section{Introduction}

The Community Engagement Working Group on Career Pipeline and Development worked on two contributed two papers below. We report findings and recommendations based on them
\begin{enumerate}
\item Facilitating Non-HEP Career Transition~\cite{CPD_1}
\item Enhancing HEP research in predominantly undergraduate institutions and community colleges~\cite{CPD_2}
\end{enumerate}

Seeking jobs after a physics degree is a natural progression at different career transition points - postdoc, PhD, undergrad. Career decisions are guided by long-term priorities in life, such as, aspirations, interests, values, economic situation, financial security and personal situations among many others. Though a career in HEP is highly desired by early career scientists, academic jobs over the decades have not kept pace with number of job seekers. Reasons to pursue a degree or particular area of research may evolve over the course of a degree at the undergraduate, graduate (Masters/PhD), postdoc or even more senior levels. On the other hand, physics degree holders typically have numerous job and career opportunities available that apply their skills outside the field. Industry job are openings new areas like  Data Science, Machine Learning and Quantum computing. Hence, physics degree holders have a greater than ever opportunity to build a career in industry. However an organised guidance to facilitate HEP to industry transition is lacking. Concurrently, faculty jobs for postdocs at R1 institutes have almost plateaued in past decades and remain highly competitive and selective. While two-thirds of physics degree holders - postdocs, PhD or undergrads seek employment in industry, postdocs or fresh PhD degree holders may also seek employment in non-R1 institutes like predominantly undergraduate institutions (PUI) and community colleges (CC). Subsequently, new faculty hires a PUIs or CCs  may decide not to pursue research but if they do, they face daunting issues like higher teaching load, lack of local research infrastructure, HEP experiment requirements like service work and funding challenges. In sections below we present key questions, findings and recommendations on these two topics based on discussions through the past two years and a community wide survey. Please note that at the time of doing survey we used the term "HEPA" (High Energy Physics and Astrophysics) but then it was decided to just use "HEP". Some figures may contain the word "HEPA".

\section{Facilitating Non-HEP Career Transitions}
HEP research gives undergraduate students, graduate (Masters or Ph.D.) students, and postdocs a wide array of scientific and technical skills that open up a a variety of career paths beyond academia. At the same time, the number of faculty and scientist positions at universities and other institutions is not keeping pace with the number of job seekers. Hence, private or government sector employment (collectively called ``industry") is the career path taken be more than two-thirds of trained physicists~\cite{nonhep_aip, nonhep_aipindustry}. Not only does industry provide the highest number of jobs~\cite{aip_afterPhD, phytoday_afterPhD,physicstoday_careers} but also the highest salaries. Early career physicists may exit academia at any stage -- from directly after the undergraduate degree to after several years of postdoctoral work. While the time spent earning a Ph.D.~in HEP is comparable to the other fields, postdocs usually last 5-6 years, a bit on higher side compared to other disciplines. Long postdoctoral periods with low salaries before taking an academic job or leaving for industry represent a significant risk for personal stability. It is therefore beneficial for all HEP trainees to engage in smart career planning, have awareness of various possibilities in industry for jobs, and network with alumni and colleagues to facilitate the process of considering jobs in industry at all stages. \\
\\
Several key questions listed below articulate these scenarios:

\begin{enumerate}
\item What are the existing efforts to facilitate transition to industry jobs?
\item What is the existing attitude/support in HEP towards industry jobs?
\item How can we strengthen networking with HEP alumni?
\item How can we be proactive in helping young career in preparation for an industry job?
\item Can we encourage alumni to contribute or come back to research and reverse brain drain, are alumni even interested?
\end{enumerate}

The Snowmass Early Career~\cite{nonhep_snowmassyoung} (SEC) team prepared and conducted a survey between June 28 and August 15, 2021, for the HEP community. The survey was designed to inform the survey team on opinions, experiences, and outlook regarding several topics; including careers, physics outlook, workplace culture, harassment, racism, visa policies, the impacts of COVID-19, and demographics. The Community Engagement Frontier (CEF) Working group (WG) on Career Pipeline and Development (CPD) strategically planned and structured its questions and contributed to this survey to collect feedback related to key questions posed above. The participants included HEP physicists (Undergraduates, Masters and PhD students, Postdocs, Engineers, Technicians, Teaching faculty, Tenure-track faculty, Tenured faculty and Scientists or Senior scientists from national lab and universities.) and HEP alumni and the input is broadly divided accordingly. 

The feedback from the survey informs several recommendations to Snowmass for measures that the HEP community should take to facilitate career transitions to industry. Feedback from HEP alumni provides a sense of the challenges in transitioning from HEP, and the measures we can take to fix the current shortcomings.  The Snowmass CPD-WG presents recommendations on professional development for industry careers, deepening connections with HEP alumni, and strengthening partnerships between HEP and industry.

\subsection{Professional development for industry careers}

A professorship comes with its lifetime of studying science, and with the very strong appeal of a comfortable intellectual life devoted to understanding nature at its most fundamental level. However, the path to a professorship comes with many challenges and uncertainties in terms of work/life balance, pressure to publish, competition with scientists across the globe for a limited number of jobs, limited choice of geographic location, pressure to secure grants, and lesser material compensation compared to the industry. At the same time, many fulfilling career paths exist for physicists who are willing to embrace a transition to industry. For some, what is sought is an industry job that is sufficiently interesting and exciting, so as to allow the use of current skill set in a creative way, as well as challenging enough to foster the development of new skills and allow contributions to have a positive impact in the world. It may be hard to get "out of the box" but the only way to make sense out of "change" is to plunge into it, move with it, and join the "dance". This applies to exiting the field at any level. Industry-minded professional development opportunities along the road toward an academic career are needed to make a career transition possible and even, perhaps, exciting.

\begin{figure}[h]
\centering
\begin{minipage}[b]{0.5\textwidth}
    \includegraphics[width=1.0\hsize]{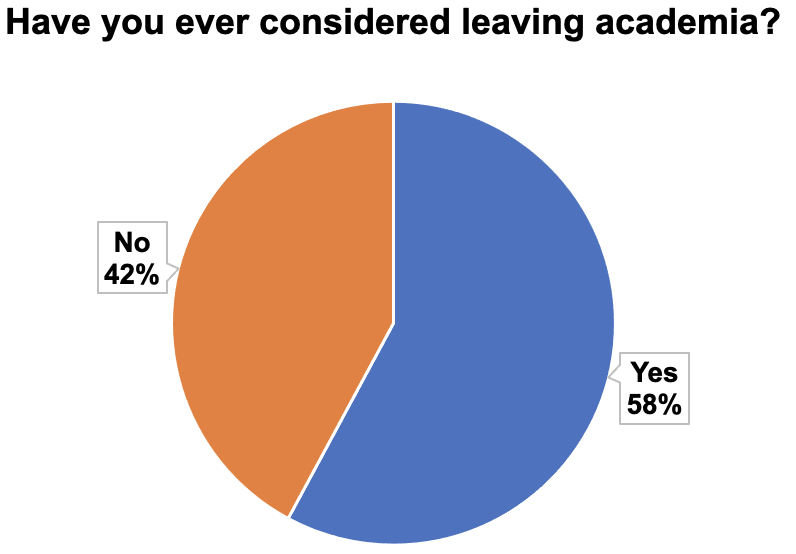}
 \caption{}
    \label{fig:nonhep_Q42}
    \end{minipage}
\end{figure}

\subsubsection*{Findings:}

\begin{figure}[h]
\centering
\includegraphics[width=1.0\hsize]{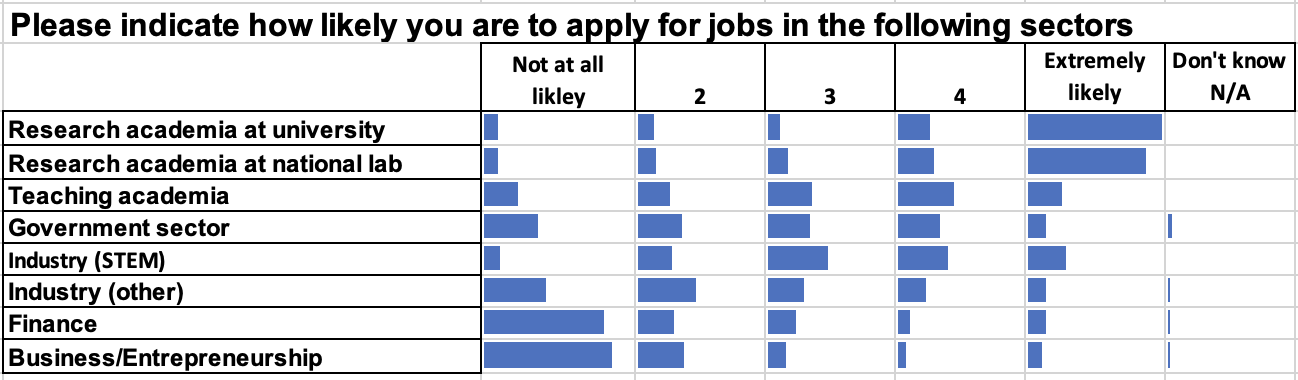}\hspace{5pc}%
\begin{minipage}[b]{28pc}\caption{\label{fig:nonhep_Q27}}
\end{minipage}
\end{figure}

\begin{figure}[h]
\centering
\includegraphics[width=1.0\hsize]{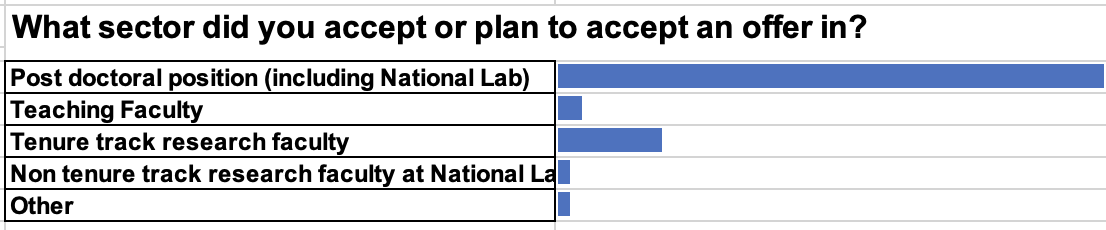}\hspace{5pc}%
\begin{minipage}[b]{28pc}\caption{\label{fig:nonhep_Q23}}
\end{minipage}
\end{figure}

\begin{figure}[h]
\centering
\includegraphics[width=1.0\hsize]{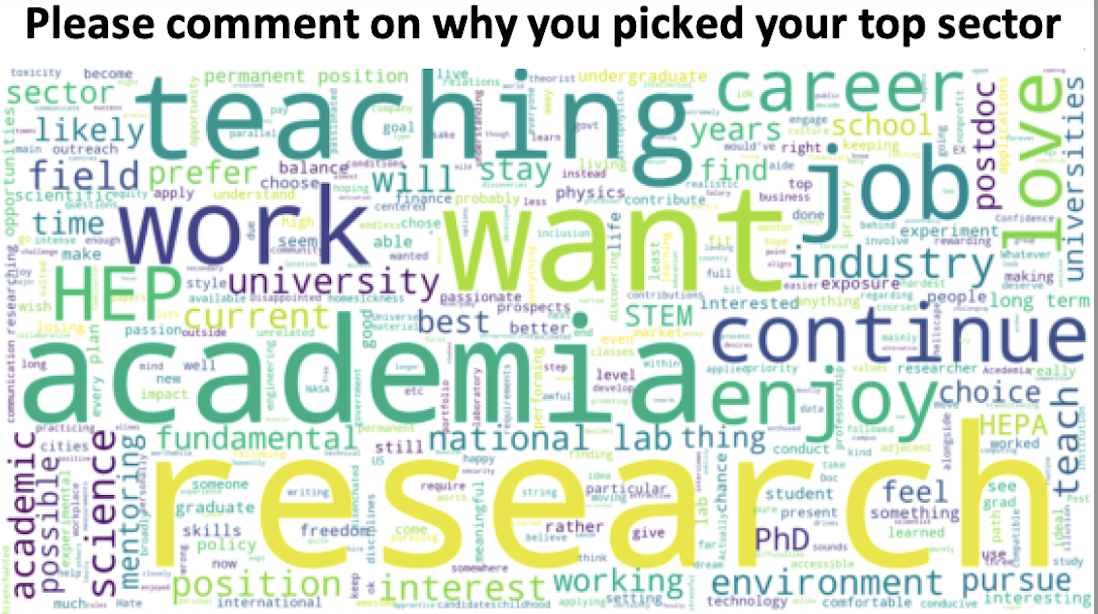}\hspace{5pc}%
\begin{minipage}[b]{28pc}\caption{\label{fig:nonhep_Q28}}
\end{minipage}
\end{figure}

In the SEC survey, almost half of the participants noted that they have considered leaving academia (Figure~\ref{fig:nonhep_Q42}), making it essential to organise and facilitate this transition. While it is rare, even faculty members (tenured or nontenured) may seek out industry jobs for personal or other reasons. But attitudes are somewhat negative about this change -- respondents' primary job preference is research in academia (university faculty) or national labs (scientist), and the least preferred job fields are business or entrepreneurship, as shown in Figure~\ref{fig:nonhep_Q27}. Most participants, being young, have accepted or plan to accept academic positions that can constitute an intermediate step in an academic career ladder; like going from PhD position to a postdoc, or from a postdoc to a  faculty position, as shown in Figure~\ref{fig:nonhep_Q23}, which again indicates a strong preference to stay in HEP. Figure~\ref{fig:nonhep_Q28} shows the reason to apply for a future academic or industry job is to continue to pursue "research". This implies that physicists' career preference is to ``do research" whether in academia or industry.


\begin{figure}[h]
\centering
\includegraphics[width=35pc]{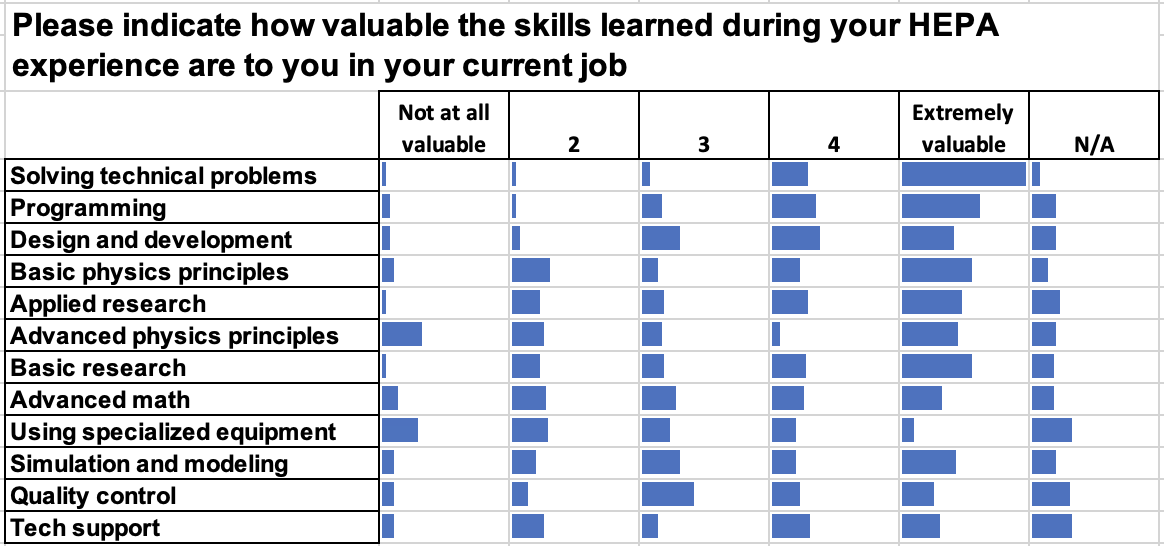}\hspace{5pc}%
\begin{minipage}[b]{28pc}\caption{\label{fig:nonHEP_Q71}}
\end{minipage}
\end{figure}

Skills acquired during physics training are central to the job landscape in industry, and offer routes to productive employment in varied and rewarding careers. Figure~\ref{fig:nonHEP_Q71} shows that HEP alumni feel that almost all HEP skills are valuable in current industry jobs. Not only technical skills are valuable -- it is not uncommon for HEP physicists is to work with non-profit organisations or as science writers~\cite{nonhep_sciencewriter1,nonhep_sciencewriter2} or on other academic and scholarship activities; for example, ~\cite{nonhep_yangyangcheng} when personal passion is a big driving factor. 

\begin{figure}[h]
\centering
\begin{minipage}[b]{0.5\textwidth}
\includegraphics[width=0.955\linewidth]{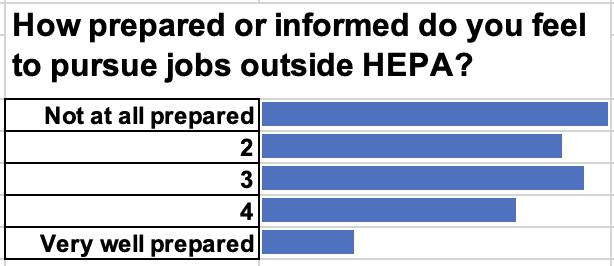}
\caption{}
\label{fig:nonhep_Q22}
\end{minipage}
\begin{minipage}[b]{0.4\textwidth}
\includegraphics[width=1.155\linewidth]{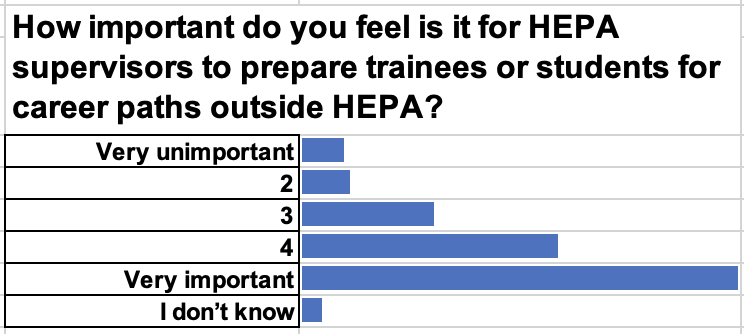}
\caption{}
\label{fig:nonhep_Q62}
\end{minipage}
\end{figure}

However, the level of unpreparedness for non-HEP jobs is overwhelming, as indicated in Figure~\ref{fig:nonhep_Q22}. There is not enough mentoring and preparedness on part of supervisors and mentors to prepare their students for career paths in industry, as shown from the survey result in Figure~\ref{fig:nonhep_Q62}. There are networking events for industry careers targeted for HEP community were set off at CERN by the CMS Collaboration~\cite{nonhep_networkCMS2013,nonhep_networkCERN2013} a decade ago and later adopted by all CERN Experiments~\cite{nonhep_networkCERN2016,nonhep_networkCERN}. It eventually led to the creation of the CERN Alumni Network~\cite{nonhep_networkCERNalumni} and the CERN Career Fairs~\cite{nonhep_networkCERNcareer}. In the USA there are some efforts in that direction; like those by the Fermilab Student Postdoc Association~\cite{nonhep_fspafermilab}. But clearly a more organised effort is required.

\subsubsection*{Recommendations for professional development:}

\begin{itemize}
    \item \textbf{Supervisors and mentors} should be directly involved in planning the career of their mentees early on. This career plan \textbf{should not be based on the desires of the mentor but the skills and interest of the mentee}. A commensurate effort in the job search process is also needed.
    
    \item \textbf{Supervisors} should allow a certain fraction of working time for their mentees to pursue opportunities and preparation activities for a possible industry career.

    \item \textbf{HEP experiments, laboratories, or university departments} should provide training for supervisors so that they can better understand and be more sensitive to the needs of their mentees in terms of their career goals and preparation. 

    \item \textbf{HEP experiments and/or laboratories} should provide workshops on industry job preparation:  translating HEP skills and examples to industry language, converting CVs to resumes suitable for different fields, finding successful job search phrases (for example, ``Engineer" or ``Data Scientist" as opposed to ``Physicist"). \textbf{This will be most successful when paired with the recommendations below for deepening connections with HEP alumni.}
    
    \item \textbf{HEP experiments and/or laboratories} should develop innovative opportunities for networking with HEP alumni in various fields to strengthen industry job search success. Alumni are more than willing and happy to respond and engage. \textbf{This will be most successful when paired with the recommendations below for deepening connections with HEP alumni.}
    
\end{itemize}

\subsection{Deepening connections with HEP alumni}

\subsubsection*{Findings:}

\begin{figure}[h]
\centering
\includegraphics[width=35pc]{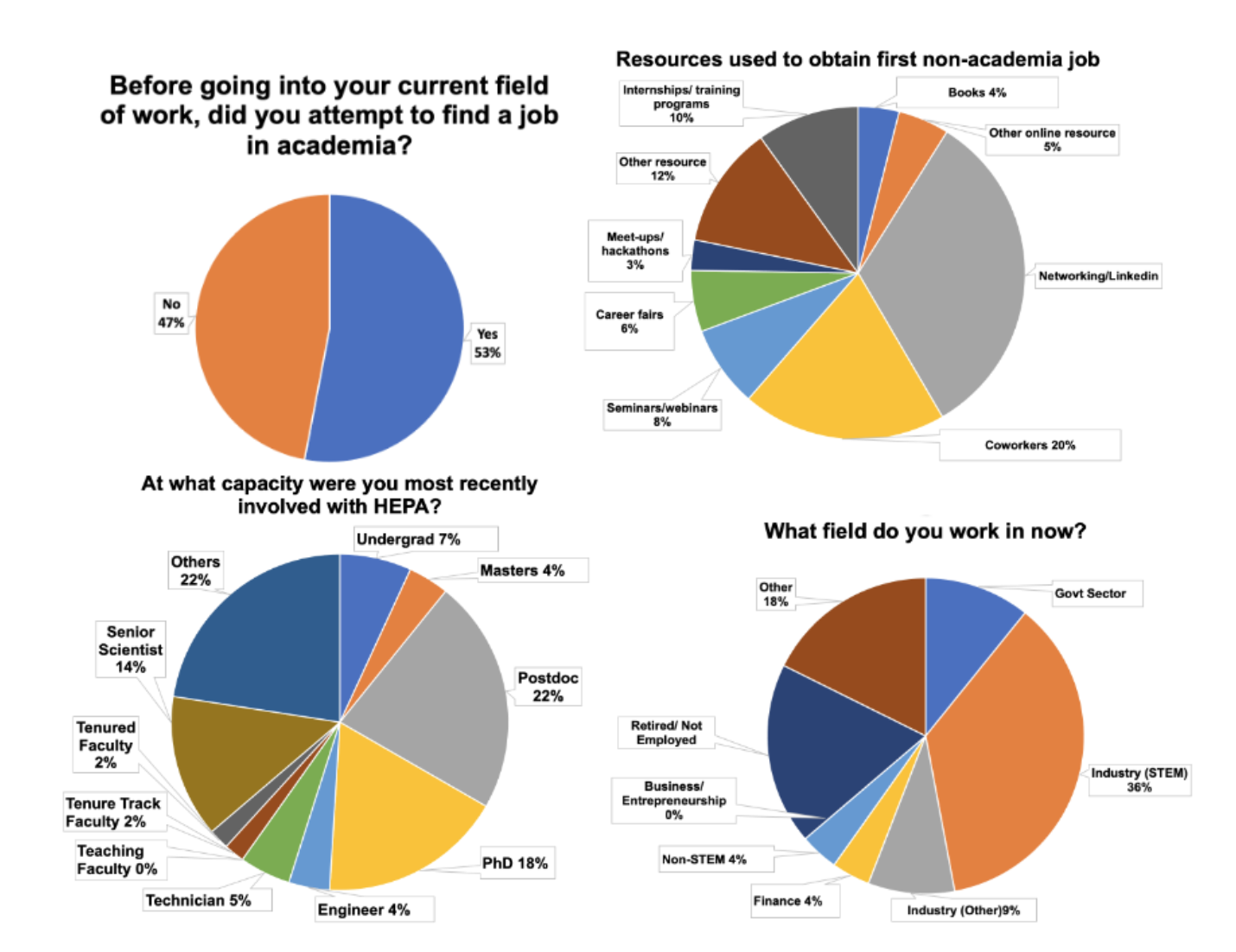}\hspace{5pc}%
\begin{minipage}[b]{28pc}\caption{\label{fig:nonhep_Q69Q70Q66Q68}}
\end{minipage}
\end{figure}

Almost half of HEP alumni attempted to find direct employment in the field, as shown in Fig.~\ref{fig:nonhep_Q69Q70Q66Q68} (top left). Failing to stay in the field reflects the shortage of permanent jobs in HEP and, simultaneously, lucrative opportunities in industry, particularly higher salaries. Figure~\ref{fig:nonhep_Q69Q70Q66Q68} (bottom left) shows the final involvement level with HEP before exiting the field -- almost 50\% of responding alumni transitioned after the completion of a student or postdoctoral position, before acquiring a faculty/scientist position. Figure~\ref{fig:nonhep_Q69Q70Q66Q68} (bottom right) shows the variety of sectors in industry in which alumni currently work, with a majority of them in STEM-related fields.  Figure~\ref{fig:nonhep_Q69Q70Q66Q68} (top right) shows that the most important resource to obtain industry jobs were networking and co-workers. 


\begin{figure}[!h]
\centering
\includegraphics[width=35pc]{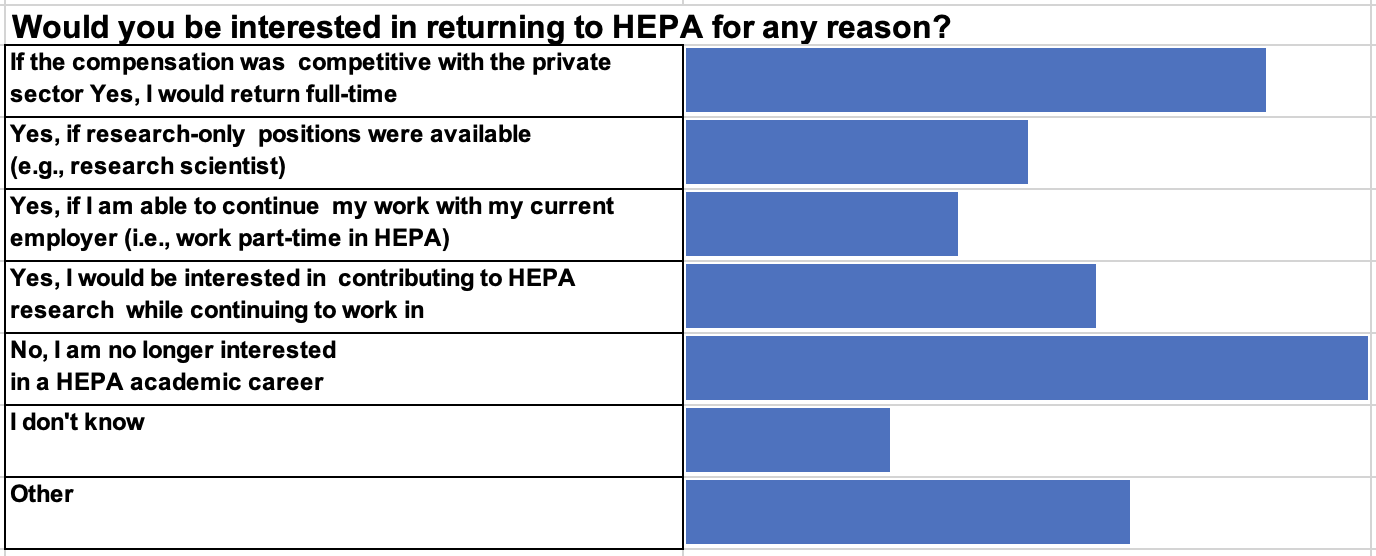}\hspace{5pc}%
\begin{minipage}[b]{28pc}\caption{\label{fig:nonhep_Q72}}
\end{minipage}
\end{figure}

\begin{figure}[!h]
\centering
\begin{minipage}[b]{28pc}
\includegraphics[width=1.0\linewidth]{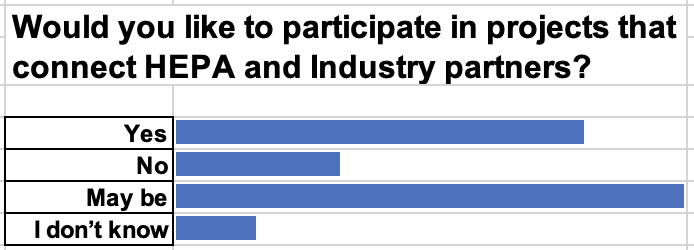}
\caption{}
\label{fig:nonhep_Q33}
\end{minipage}
\end{figure}

\begin{figure}[!h]
\centering
\begin{minipage}[b]{28pc}
\includegraphics[width=25pc]{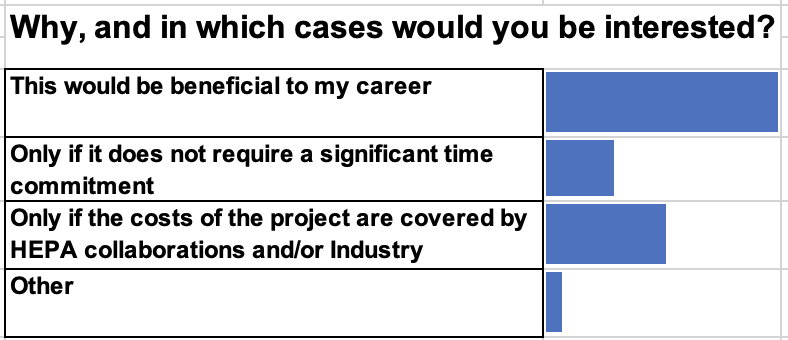}
\caption{}
\label{fig:nonhep_Q34}
\end{minipage}
\end{figure}

During the course of working in the Snowmass CPD-WG meetings over the past several months we had presentations and direct discussions with several HEP-alumni. Most of them exited the field after a PhD degree or postdoc (refer to their experiences in these presentations~\cite{nonhep_remington,nonhep_vasel,nonhep_pasner}). The most important step was the difficult decision to make up their mind to leave academia at such relatively late stages, which also connects to personal matters for many. It looks challenging for alumni to return to HEP, as indicated in Figure~\ref{fig:nonhep_Q72}, but large fraction are willing to participate in projects connecting HEP and industry (Figure~\ref{fig:nonhep_Q33}), and many see value in doing so (Figure~\ref{fig:nonhep_Q34}). 

\subsubsection*{Recommendations for alumni connection:}

\begin{itemize}
    \item \textbf{Supervisors and mentors} should actively communicate with alumni and highlight their experiences for current students and postdocs, to normalize the reality of transitioning to an industry career.
    
    \item \textbf{The US HEP community should develop tools and portals for connecting with alumni}. Existing programs for networking with alumni like at CERN must be studied and adopted. This effort should be supported and strengthened by funding agencies by dedicating a small amount of continuous funding to support technical and personnel staff that can organise and build framework that can serve as a hub to facilitate process of networking with alumni. \textbf{A DOE lab would be an ideal place to host this effort}, like Fermilab, which is a hub for US particle physics.
    
    \item \textbf{HEP experiments and laboratories} should take creative steps to reverse ``brain drain" from HEP by exploring mechanisms for collaboration with alumni on HEP projects. 
    \begin{itemize}
        \item Alumni are a relatively low cost but very valuable asset with an abundance of experience from transitioning to an industry career. Their goodwill to contribute and strengthen ties with HEP can be tapped to facilitate industry job transitions and further the goals of both groups.
        \item Individual scientific collaboration can be extended to the company of the alumni itself and this can strengthen knowledge transfer from labs and universities and vice versa; and work done by HEP research can benefit companies and vice versa. A recent example has been the use of Amazon Web Service (AWS) with the CMS experiment workload through HEP Cloud project~\cite{nonhep_aws}.
    \end{itemize}  
\end{itemize}

\subsection{Strengthening industry partnerships}

\subsubsection*{Findings:}

\begin{figure}[h]
\centering
\includegraphics[width=30pc]{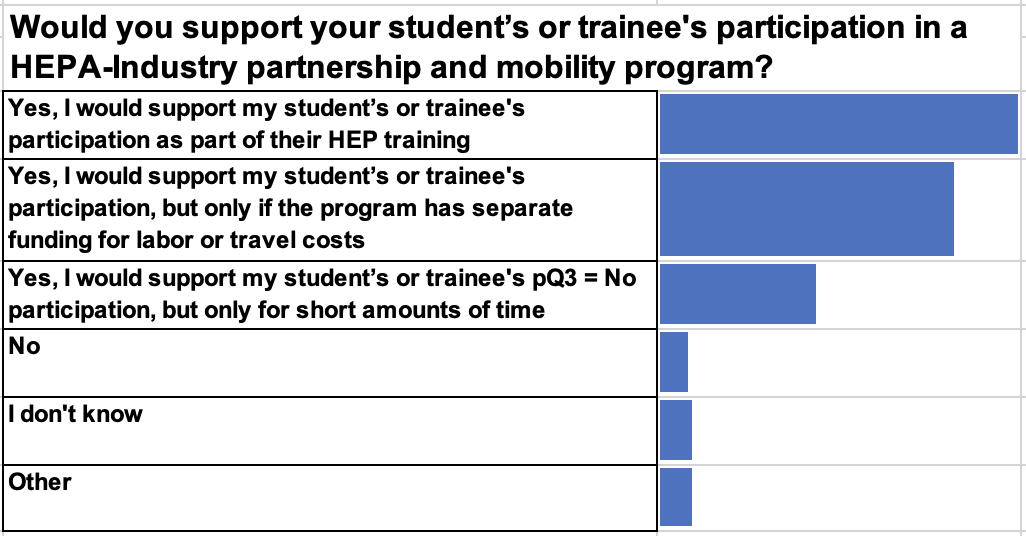}\hspace{5pc}%
\begin{minipage}[b]{28pc}\caption{\label{fig:nonhep_Q64}}
\end{minipage}
\end{figure}

Among supervisors, there is a strong support for their students and mentees to participate in HEP-industry partnerships and mobility programs that would prepare them well for industry jobs, as shown in Figure~\ref{fig:nonhep_Q64}. However, it is also shown that funding for such partnership activities must come from a source separate from their core HEP research support.

National laboratories could offer internships or training programs for graduate students in the areas of Accelerator Technology, Computer and Information Science, Detector and Engineering Technology, Environmental Safety and Health and Radiation Therapies to facilitate the training process for a career in industry. To go further, partnerships could offer HEP students or scientists the opportunity to work directly in an industry setting. While there are many examples of internships sponsored by the National Labs and funding agencies, like DOE and NSF, at the labs or universities, there are significantly fewer internship opportunities for HEP early career scientists or students in industry. Industry internships can serve as a very good opportunity to experience industry culture and environment and apply HEP skills. This would require students time off from core research to allow a period of time to focus on industry internship, but this needs careful planning and should not dilute the focus on core research goals.

\subsubsection*{Recommendations for industry partnerships:}
\begin{itemize}
    \item \textbf{HEP laboratories} should create targeted internships or training programs in the areas of Accelerator Technology, Computer and Information Science, Detector and Engineering Technology, Environmental Safety and Health and Radiation Therapies. This would expand access to industry-focused training to students and postdocs who are not based at national laboratories.
    \item \textbf{HEP laboratories} should leverage existing public-private partnerships~\cite{nonhep_fnalindustry,nonhep_cernindustry} with industries like Accelerator Technologies, Computers Information Science, Detector and Engineering Technologies and also Environmental Safety to create experience for resident students and early career scientists to build skills ad connections for a future industry career.
    \item \textbf{Funding agencies} should evaluate funding rules and regulations to allow HEP students and postdocs to pursue industry-focused training that can be integrated with their core research curriculum.
    \item \textbf{Supervisors} must adopt a mindset that industry partnerships and career transitions are valuable options for their students and postdocs, and should support their participation in training opportunities whenever possible. 
\end{itemize}



\section{Enhancing HEP research in predominantly undergraduate institutions and community colleges}

The long-term success of HEP lies in expanding inclusiveness beyond national labs and academic research institutions to a vast community of predominantly undergraduate institutions (PUI) and community colleges (CC). Institutions such as PUIs and CCs offer an early starting point in the career pipeline that can mitigate issues of lack of diversity and underrepresented participation of different groups in HEP. However, there are many underlying systemic, structural, and cultural challenges that need to be addressed collectively. Experimental collaborations are largely populated by national labs and research-focused academic institutions (non-PUIs). The faculty at PUIs and CCs have a high teaching load that is detrimental to their research participation. In addition, there is a lack of guidance, access, and tough competition for securing research funding. The students also suffer from a lack of research infrastructure and technical equipment that can only be found at national labs and larger universities. There are existing successful efforts to enhance the HEP research experience of students and faculty members that can be leverage to provide more research opportunities and establish a sustainable national program targeting specifically the issues faced by communities at PUIs and CCs. Enhancing research support, mentoring and skill building for these faculty members and their students would broaden the spectrum of population in HEP and impact the scientific workforce preparation of our society.

Several key questions on this topic are given below: 

\begin{enumerate}
\item Can cutting-edge research be even done at a PUI or CC, given the high teaching load and the expertise disparity between undergrads and grad students?
\item In a highly collaborative field, can single faculty at PUIs or CCs really make a contribution?
\item PUIs and CCs engage a {\it fundamentally different} student population and often provide a very different set of experiences for the students when compared to larger research institutions. Is this argument strong enough to direct some fraction of our limited funding resources to these institutions? 
\item Is it important for any scientific field to reflect the broader society in which it operates? If the field does not reflect from that society, does it risk losing their support?
\end{enumerate}

\subsection{Institutional Culture}

While almost all HEP students are enrolled at major research universities, 40\% of undergraduate students in the United States are currently enrolled in a CC. A large proportion of those students are from demographics typically underrepresented in STEM fields [15, 16]. 
Nearly 80\% of CC students indicate their goal is to earn a bachelor’s degree [16], but they face many challenges that unfortunately lead to nearly 70\% of students dropping out before completing their degree [17], such as working or caring for family. 
These non-traditional students can still be just as passionate about science as their peers at other colleges and universities and should have access to teachers who are engaged with scientific research. 

Students attending PUIs have chosen a small-school experience for their undergraduate degree, perhaps due to a racial or religious connection with the university. Physics is often interdisciplinary with engineering and other sciences at PUIs, so a more diverse pool of students could gain exposure to HEP research. At PUIs, students may be at a ``pivot-point" in their lives as they decide to pursue grad school -- connection with professors engaging in research can be life-changing for them. If research opportunities are confined to larger research institutions, we are potentially excluding, or at least minimizing, a particular cross section of individuals. PUIs provide a prime opportunity to offer research experiences that can mitigate these leaky pipelines where we might lose the talent of women, persons of color, students with disabilities, or other under-represented minorities. Faculty at PUIs are uniquely positioned to reach out to individual students in an informed manner and are often trained to engage with DEI issues, LGBT+ and Ally training, or guidance in dealing with mental health issues. 

\subsubsection*{Findings:}

Faculty at PUIs and CCs are fighting a perception in HEP collaborations that they cannot make valuable contributions to the field from their ``non-research" positions. However, because faculty at PUIs are incentivized to stay current with educational research and best-practices they often have something to offer their collaborators, particularly in helping train graduate students and post-docs to be better teachers. Participation by PUIs is enriching to research collaborations, and can help to change the perception that a preparation for a faculty job only requires being a good researcher -- it also requires being a good teacher! 

The PUI or CC faculty who do participate in HEP research face a lack of understanding among their colleagues and university administration about the requirements and regulations of working on an experiment that is hosted at an external laboratory (which may be outside the country). Dedicated communication from experiments would help PUI administrators appreciate the benefit of having faculty participate in research activities with large, external collaborations, as those faculty members seek tenure, promotion, or support for funding proposals.
    
\subsubsection*{Recommendations for institutional culture:}

\begin{itemize}
    \item \textbf{The HEP community} should encourage a global shift in perception, acknowledging that:
    \begin{itemize}
        \item undergraduate research experiences are {\it key} to engaging a broader section of the student population.
        \item PUI or CC faculty have much to offer their collaborations, particularly in experiment-wide training and educational activities.
    \end{itemize}
    .  
    \item \textbf{HEP experiments} should offer coordinated communication from leadership to PUI administrators, 
    extolling the features of high energy physics research alongside highlighted participation. 
    
    \item \textbf{The HEP community} should offer special sessions for PUI and CC faculty at national meetings to develop a deeper sense of community. 
\end{itemize}

\subsection{Research Funding}

Given the lack of non-PUI faculty positions, many postdocs are seeking jobs at PUIs that inherently offer potential to tap a diverse pool of candidates for the future STEM work force. Challenges include very high teaching load and lack of lab infrastructure. The lack of start-up funding opportunities limits practical research participation. 

\subsubsection*{Findings:}

Typical funding agency awards for HEP research do not permit ``course buyout" in the budget proposal. For faculty at PUIs, changing this policy would be a big boost for research productivity and would help the funding agencies create a larger and more diverse STEM workforce.

    
More funding for ``visiting scientist" programs would especially help those PUI and CC faculty without regular funding engage in research along with their students (perhaps enabling success in future funding proposals).
Experiments can also directly offer summer research internships at member non-PUIs and national labs targeting students from under-represented groups who otherwise might not see themselves engaging in this research and might also not be able to compete for limited national undergraduate research experiences.

The key constituency for expanding the HEP workforce at PUIs are current postdoctoral researchers. If they take a non-PUI faculty position, grant writing is typically undertaken with the support of a large and established HEP group, which at a PUI they will write a proposal alone. Federal funding for programs that support grant-writing workshops would help increase participation in HEP research by preparing more postdocs to write a single-PI proposal. Under such programs, researchers who have been successful in securing grants will be paired with students and early-career professionals, creating a network of professionals invested in broadening access to HEP grants. Such a program can be modelled after the Broadening Experiences in Scientific Training (BEST) - an initiative of the National Institutes of Health (NIH) and  a program originally developed to strengthen the biomedical research workforce. To support this initiative, US experimental collaborations should ensure that postdocs are trained in typical budgeting and funding procedures for institutions in their collaboration. 

\subsubsection*{Recommendations for research funding:}

\begin{itemize}

    \item \textbf{Funding agencies} should strengthen participation by PUIs in HEP by allocating funds for grants from these institutions, and \textbf{HEP experiments or laboratories} should fund grant-writing workshops. 
    
    \item \textbf{Funding agencies} should allow course buyouts in proposals by PUI/CC faculty in order to boost productivity and establish continuity in PUI research programs.
        
    \item \textbf{Funding agencies, HEP experiments, and laboratories} should create or support paid summer programs for PUI faculty to work at National Labs or non-PUIs, as well as research opportunitites for students not enrolled at major HEP institutions.
    
    \item \textbf{Supervisors and HEP experiments} should provide training to interested students and postdocs on US-specific research funding procedures. 
    
\end{itemize} 

\subsection{Experiment Participation}

In terms of practical collaboration in a large experiment, PUI faculty are at a big disadvantage compared to the non-PUI institutions that form large fraction of HEP collaborations. The expected workload of physics analysis, service work, and future experiment development is taken on by one faculty member, who typically has a much higher teaching load that non-PUI collaborators. Experiments do not provide specific policies to mitigate these challenges faced by small institutions.

\subsubsection*{Findings:}

\begin{figure}[h]
\centering
\includegraphics[width=35pc]{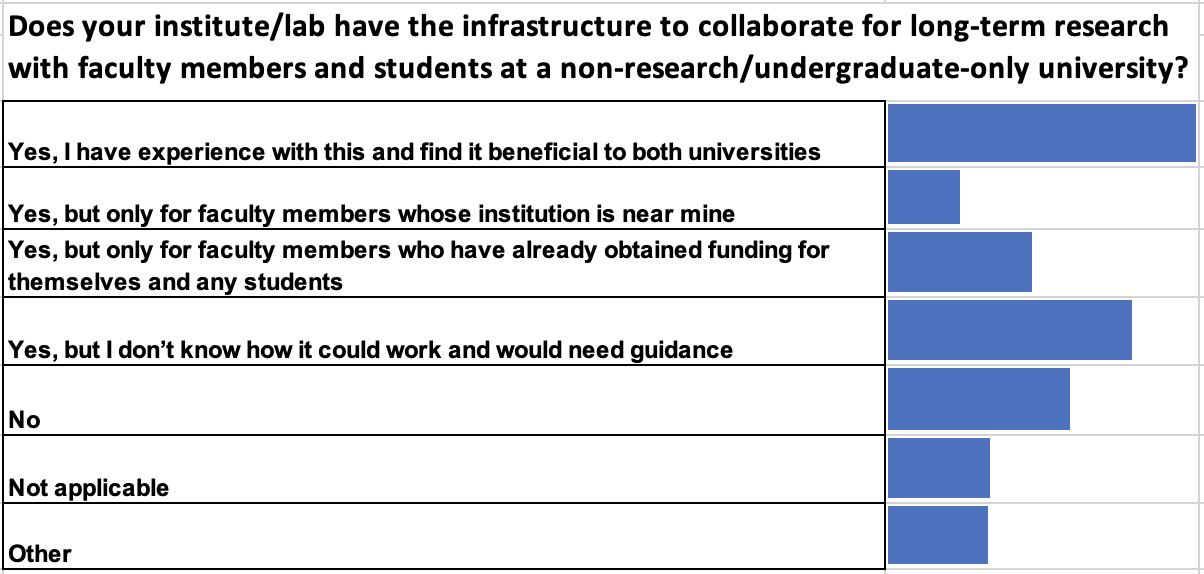}\hspace{5pc}%
\begin{minipage}[b]{28pc}\caption{\label{fig:nonhep_Q61}}
\end{minipage}
\end{figure}

For PUIs to succeed in HEP experiments, non-PUIs on HEP experiments should closely work with PUIs. This mitigates the problem of PUI faculty needing a "home institute" that is a member of the experiment and provides access to experiment's infrastructure like computing accounts. In addition, PUI students get access to postdocs, scientists and Ph.D.~students at the non-PUI as additional mentors. There is a strong  support for collaboration between research and non-research/undergraduate institutions as shown in Figure~\ref{fig:nonhep_Q61}. While it may be difficult for PUIs to eventually become independent members of a collaboration, pathways should be explored by leadership. 

Financial resources are a major limitation for PUIs seeking to perform HEP research. Some large experiments impose an entry fee and an annual fee for authorship. An entry fee structure that scales based on the number of Ph.D.~holders in the institution would be more equitable and would not disproportionately burden PUIs that lack resources compared to non-PUIs. One option might be a ``single PI" membership, or a membership that allows access to the experiment and laboratory facilities to faculty and undergraduate students without full authorship on all collaboration papers. 

Faculty members at PUIs often have much less travel support than non-PUI peers, so presence at the host laboratory is difficult to maintain. When full membership in a collaboration requires a certain quota of shift work per institution, small groups at PUIs are at a significant disadvantage because travel to a host laboratory is typically not possible during the academic year. Finally, PUI faculty typically supervise their students and perform experimental service work personally, without postdoctoral support. 

\subsubsection*{Recommendations for experimental participation:}

\begin{itemize}
    \item \textbf{Non-PUI senior-level researchers} should investigate how their groups could offer opportunities for short-term and long-term collaboration on their experiment to faculty and/or students at local PUIs. 

    \item \textbf{HEP experiments} must reevaluate large fixed ``entry fees" per institution, if they exist. Consider implementing ``light" membership forms that are low cost but not time limited. 
    
    \item \textbf{US HEP experiment leaders} should advocate with international experiment leadership for pathways to sustainable membership for PUIs, which are most common in the US. Postdocs should be aware of options for entering these pathways so they are not discouraged from applying to PUI faculty positions.
    
    
    \item \textbf{HEP experiments} must continue to improve options for remote participation in experiment meetings and service tasks, especially operational shift work.  
\end{itemize}

\section{Interconnections and Synergy with Frontiers and Topical Groups}

\textbf{CommF4, the Topical Group on Physics Education}: Physics Education is central to creating skilled workforce pipeline to all HEP Frontiers and beyond for STEM areas in Industry. All recommendations in CommF4 are strongly endorsed by this Working Group. Beyond regular course curriculum, Software Training programs and Open Science activities~\cite{PE_14} can go a long way attracting talent in HEP as well as preparing HEP talent for STEM industry.\\

\textbf{AF1, the Topical Group on Beam Physics and Accelerator Education}~\cite{nonhep_af1}: Some HEP experimentalist physicists transition to accelerator physics that is central to accelerator complex across several national labs and beyond. Applications of accelerator science in medical therapies, photon and particle probes in industry, material science, chemistry, biology, pharmaceutical development and applied nuclear science continue to stimulate demand for the expertise of well-trained specialists in accelerator science and technology and accelerator related software.\\

\textbf{NF05/10, the Topical Groups, respectively, on Neutrino properties and Neutrino Detectors and other smaller experiments}~\cite{nonhep_nf5nf10}: Small scale experiments like Rare Processes/Neutrino/Dark Matter/Nuclear Physics, can provide a healthy breeding and training ground for careers of PhD and postdoc researchers, by providing a unique set of diverse opportunities. It can serve as a bridge to a gateway of career options in academia or industry. Due to the typical short span of such experiments young members can take ownership of significant aspects of a project throughout their tenure, and contribute significantly to multiple aspects of an HEP experiment like the design, construction, operations, and data analysis. This enables early career scientists from these experiments to enter the job market with a strong and transparent portfolio, in terms of publications and contributions that may tend to get obscure in big experiments, especially when applying for jobs within or outside HEP.\\

\textbf{IF/AF6 and CompF3, the Topical Groups, respectively, on Instrumentation Frontier, Advanced Accelerator Concepts and Machine Learning}~\cite{nonhep_if2,nonhep_af6,nonhep_commf3}: Training in Detector and Instrumentation Technologies and engagement with the industry in applications and technology transfers from the particle physics community can greatly enhance career transition to a related industry. Schools like  ISOTDAQ~\cite{nonhep_isotdaq}, EDIT~\cite{nonhep_edit} and ESHEP~\cite{nonhep_eshep} and the CERN/FNAL Collider Schools~\cite{nonhep_colliderschool} can stimulate young careers that can fuel innovation not only in HEP but also in  industry. Machine Learning skills acquired to study patterns in HEP data has a great demand in data science industry.

\section{Conclusions}
To shape HEP into a compelling program of scientific discovery that spins talented STEM workforce into academia and industry it is very critical that we adopt and enable methods that lead to diverse, equal opportunity and inclusiveness from the very onset. Faculties at PUIs and CCs who provide us access to a big spectrum of diverse students must be supported to pursue research in every way possible by the funding agencies as well as by the HEP experiments. An opportunity to empower them must not be missed for a better and strong future of HEP. 

To transition skills learnt in HEP research into industry jobs, where more two-third of physic degree holders find employment, a more broad and systematic approach to the marketing of career opportunities to early career physicists is needed. While organizations such as the APS and national lab user groups can tailor and offer programs that meet their specific goals, a bigger effort is required if optimal outcomes at both the structural and personal levels are to be achieved for HEP  as well as the wider society. However, establishing and institutionalizing these approaches, as well as socializing them with members of the HEP community from the most junior to the most senior levels, will require both dedicated effort from the community along with relatively modest funding in order to bring early and mid-to-late career scientists together in environments and settings conducive to formal and informal networking and the building of relationships that will last a lifetime.

\bibliographystyle{utphysmod}
\typeout{} 
\bibliography{main.bib}

\providecommand{\href}[2]{#2}\begingroup\raggedright\begin{thebibliography}{10}

\bibitem{CPD_1}
S.~Malik, {\em et al.}, {\em {Facilitating Non-HEP Career Transition}},
  {\ttfamily \href{https://arxiv.org/abs/2203.11665}{arXiv:2203.11665}}.

\bibitem{CPD_2}
M.~Bellis, {\em et al.}, {\em {Enhancing HEP research in predominantly
  undergraduate institutions and community colleges}}, {\ttfamily
  \href{https://arxiv.org/abs/2203.11662}{arXiv:2203.11662}}.

\bibitem{nonhep_aip}
{\em {American Institute of Physics}}, \url{https://www.aip.org/}, 2018.

\bibitem{nonhep_aipindustry}
{\em {Physics PhDs Ten Years Later: Movement across Job Sectors}},
  \url{https://www.aip.org/sites/default/files/phd+10-jobsectormovemnt.pdf},
  2018.

\bibitem{aip_afterPhD}
{\em {EMPLOYMENT \& CAREERS IN PHYSICS}},
  \url{https://www.aip.org/statistics/reports/employment-and-careers-physics},
  2016.

\bibitem{phytoday_afterPhD}
{\em {Where do new PhDs work?}}, \url{https://doi.org/10.1063/PT.3.4591}, 2016.

\bibitem{physicstoday_careers}
{\em {Preparing physics students for 21st-century careers}}, \url{
  https://physicstoday.scitation.org/doi/10.1063/PT.3.3763}, 2017.

\bibitem{nonhep_snowmassyoung}
{\em {Snomass2021 Early Career}}, \url{https://snowmass21.org/start/young},
  2020.

\bibitem{nonhep_sciencewriter1}
{\em {Science News: Emily Conover}},
  \url{https://www.sciencenews.org/author/emily-conover}, 2022.

\bibitem{nonhep_sciencewriter2}
{\em {AAAS: Katrina Miller}},
  \url{https://www.aaas.org/programs/mass-media-fellowship/katrina-miller},
  2022.

\bibitem{nonhep_yangyangcheng}
{\em {Several article by Dr. Yangyang Cheng in The Guardian}},
  \url{https://www.theguardian.com/profile/yangyang-cheng}, 2021.

\bibitem{nonhep_networkCMS2013}
{\em {NETWORKING EVENT FOR CMS ALUMNI, STUDENTS, AND POSTDOC}},
  \url{https://indico.cern.ch/event/280839/ }, 2013.

\bibitem{nonhep_networkCERN2013}
{\em {NETWORKING EVENT FOR CMS ALUMNI, STUDENTS, AND POSTDOC}},
  \url{https://cms.cern/index.php/news/networking-event-cms-alumni-students-and-postdocs},
  2013.

\bibitem{nonhep_networkCERN2016}
{\em {The 4th ALICE, ATLAS, CMS and LHCb Career Networking Event}},
  \url{https://home.cern/news/news/experiments/4th-alice-atlas-cms-and-lhcb-career-networking-event},
  2016.

\bibitem{nonhep_networkCERN}
{\em {Events for the CERN-based community}},
  \url{https://indico.cern.ch/category/12583/}, 2022.

\bibitem{nonhep_networkCERNalumni}
{\em {The CERN Alumni Network}}, \url{https://alumni.cern/}, 2022.

\bibitem{nonhep_networkCERNcareer}
{\em {CERN @ Career fairs}}, \url{https://careers.cern/cern-career-fairs},
  2022.

\bibitem{nonhep_fspafermilab}
{\em {Fermilab Student \& Postdoc Association (FSPA): Career Activities}},
  \url{https://fspa.fnal.gov/career-activities/}, 2022.

\bibitem{nonhep_remington}
{\em {Transitioning to Industry}}, \url{
  https://indico.fnal.gov/event/44789/contributions/193952/attachments/132660/163282/TransitioningToIndustry.pdf},
  2020.

\bibitem{nonhep_vasel}
{\em {Application for the TNSF Fellowship: Justin Vasel}}, \url{
  https://indico.fnal.gov/event/43427/contributions/192519/attachments/131958/161806/Vasel-CommF2-TNSF-process.pdf},
  2020.

\bibitem{nonhep_pasner}
{\em {2020 Applying to the AAAS Fellowship: Jacob M. Pasner}}, \url{
  https://indico.fnal.gov/event/43427/contributions/192518/attachments/132000/161892/Pasner_SnowMass_AAAS_Presentation_July_27_2020.pdf},
  2020.

\bibitem{nonhep_aws}
{\em {Experiment that Discovered the Higgs Boson Uses AWS to Probe Nature}},
  \url{https://aws.amazon.com/blogs/aws/experiment-that-discovered-the-higgs-boson-uses-aws-to-probe-nature/},
  2016.

\bibitem{nonhep_fnalindustry}
{\em {Fermilab:Partnerships and Technology Transfer}},
  \url{https://partnerships.fnal.gov/}, 2022.

\bibitem{nonhep_cernindustry}
{\em {CERN:Partnerships}}, \url{https://home.cern/partnerships}, 2022.

\bibitem{PE_14}
S.~Malik, {\em et al.}, {\em {Broadening the scope of Education, Career and
  Open Science in HEP}}, {\ttfamily
  \href{https://arxiv.org/abs/2203-08809}{arXiv:2203-08809}}.

\bibitem{nonhep_af1}
{\em {AF1: Beam Physics and Accelerator Education}},
  \url{https://snowmass21.org/accelerator/outreach/start}, 2020.

\bibitem{nonhep_nf5nf10}
{\em {FRONTIERS IN NEUTRINO PHYSICS}},
  \url{https://snowmass21.org/neutrino/start}, 2020.

\bibitem{nonhep_if2}
{\em {INSTRUMENTATION FRONTIER}},
  \url{https://snowmass21.org/instrumentation/start}, 2020.

\bibitem{nonhep_af6}
{\em {AF6: Advanced AcceleratorConcepts}},
  \url{https://snowmass21.org/accelerator/advanced/start}, 2020.

\bibitem{nonhep_commf3}
{\em {CompF3: Machine Learning}},
  \url{https://snowmass21.org/computational/machine_learning}, 2020.

\bibitem{nonhep_isotdaq}
{\em {ISOTDAQ}}, \url{https://home.cern/tags/isotdaq}, 2020.

\bibitem{nonhep_edit}
{\em {Excellence in Detectors and Instrumentation Technologies}},
  \url{http://detectors-school.web.cern.ch/}, 2020.

\bibitem{nonhep_eshep}
{\em {The European Schools of High-Energy Physics}},
  \url{https://physicschool.web.cern.ch/eshep/default.html}, 2020.

\bibitem{nonhep_colliderschool}
{\em {The CERN-Fermilab Hadron Collider Physics Summer Schools}},
  \url{http://hcpss.web.cern.ch/}, 2022.

\end{thebibliography}\endgroup


\end{document}